\documentstyle[12pt]{article}
\renewcommand{\sectionmark}[1]{}
\begin{document}
\markboth{ }{\hfil PARAMETRICAL NEURAL NETWORK}
\centerline{\Large{\bf Parametrical Neural Network}}
\vskip 2mm
\centerline{\large Boris V. Kryzhanovsky, Leonid B. Litinskii, 
Andrey L. Mikaelyan}
\vskip 2mm
\centerline{\large{\it Institute of Optical Neural Technologies Russian Academy 
of Sciences}}
\centerline{\large{\it Moscow, Russia}}
\vskip 10mm
{\bf Acknowledgments.}

The work was supported by Russian Basic Research Foundation (grants 02-
01-00457 and  01-01-00090) and the 
program "Intellectual Computer Systems" (the project 2.45).
The authors are grateful to Dr. Inna Kaganova for preparation of this 
manuscript. 
\vskip 10mm
\noindent
Leonid B. Litinskii\\ 
\vskip 2mm
\noindent
Russia, 119333, Moscow, 
Institute of Optical Neural Technologies\\ Russian Academy of Sciences,
Vavilov Street, 44/2
\vskip 2mm
\noindent
Tel.:+7-/095/-135-1351, fax: +7-/095/-135-1351,\\
\vskip 2mm
\noindent
e-mail: litin@mail.ru\\
\vskip 2mm
\noindent
PARAMETRICAL NEURAL NETWORK

\newpage
\centerline{\large{\bf Parametrical Neural Network}}
\begin{abstract}
The storage capacity of the Hopfield model is about 15\% of the network size.
It can be increased significantly in the Potts-glass model of the associative
memory only. In this model neurons can be in more than two different states. We
show that even greater storage capacity can be achieved in {\it the parametrical
neural network} (PNN) that is based on the parametrical four-wave mixing process 
that is well-known in nonlinear optics. We
present a uniform formalism allowing us to describe both PNN and the
Potts-glass associative memory. To estimate the storage capacity we use the
Chebyshev-Chernov statistical technique.  

\end{abstract}
\newpage
\section{INTRODUCTION}
In refs. \cite{Kry02},\cite{Fon01} a network based on the parametrical 
four-wave mixing process (FWM) \cite{Blo66} that is well-known in 
nonlinear optics was 
examined. Such a network is capable to hold and handle information that 
is encoded in the form of the phase-frequency modulation. In the network 
the signals propagate along interconnections in the form of 
quasi-monochromatic pulses at $q$ different frequencies 
$$\{ \omega _l \}_1^q  \equiv \{ \omega _1 ,\omega _2 ,...,\omega _q \}.
\eqno(1)$$
The model is based on a parametrical neuron that is a cubic nonlinear 
element capable to transform and generate frequencies in the parametrical FWM-
processes
$\omega_i  - \omega _j  + \omega _k  \to \omega _r$.
Schematically this model of a neuron can be assumed as a device that is
composed of a summator of input signals, a set of  $q$ ideal frequency 
filters
$\{ \omega _l \} ^q$, a block comparing the amplitudes of the signals 
and $q$ generators of quasi-monochromatic signals $\{ \omega _l \} ^q$.

Let $\{K^{(\mu)}\}_1^p$ be a set of patterns each of which is a set of 
quasi-monochromatic pulses with frequencies defined by Eq.(1) and 
amplitudes equal to
$\pm 1$: 
$$K^{(\mu)}=(\kappa_1^{(\mu)},\ldots,\kappa_N^{(\mu)}),
\mbox{ where }
\kappa_i^{(\mu)}=\pm\exp(\imath\omega_{l_i^{(\mu)}}t),\quad
\left\{\begin{array}{l}\mu=1,\ldots,p;\\i=1,\ldots,N;\\
1\le l_i^{(\mu)}\le q.\end{array}\right.\eqno(2)$$
The memory of the network is localized in interconnections $T_{ij},\ 
i,j=1,\ldots,N$, which accumulate the information about the states of
$i$th and $j$th neurons in all the $p$ patterns. We suppose that the 
interconnections are dynamic ones and that they are organized according to
the Hebb rule:
$$T_{ij}=(1-\delta_{ij})\sum_{\mu=1}^p\kappa_i^{(\mu)}\kappa_j^{(\mu)*},
\quad i,j=1,\ldots,N.\eqno(3)$$
The network operates as follows. A quasi-monochromatic pulse with a frequency
$\omega_{l_j}$ that is propagating along the $(ij)$-th interconnection 
from the
$j$th neuron to the $i$th one, takes part in FWM-processes with the pulses 
stored 
in the interconnection,
$$\omega_{l_i^{(\mu)}}-
\omega_{l_j^{(\mu)}}+\omega_{l_j}\to\{\omega_l\}_1^q.$$
The amplitudes $\pm 1$ have to be multiplied. Summing up the results of 
these partial transformations over all patterns, $\mu=1,\ldots,p$, we obtain a 
packet 
of quasi-monochromatic pulses, where all the frequencies from the set 
(1) are present.
The amplitudes of the pulses are determined by the interconnection. This 
packet is the result of transformation of the pulse $\omega_{l_j}$ by 
the interconnection $T_{ij}$, and it comes to the $i$th neuron. All such 
packets are summarized in this neuron. The summarized signal propagates 
through $q$ parallel ideal frequency filters. The output signals from the 
filters are compared with respect to their amplitudes. The signal with 
the maximal amplitude activates the $i$-th neuron  ('winner-take-all'). As a 
result it 
generates an output signal whose frequency and phase are the same as the 
frequency and the phase of the activating signal. 

Generally, when three pulses interact, under a FWM-process always the 
fourth pulse appears. The frequency of this pulse is defined by the 
conservation laws only. However, in order that the abovementioned model works 
as a memory, an important condition must be add, which has to facilitate 
the propagation of the useful signal, and, in the same time, to 
suppress external noise. This condition is {\it the principle of 
incommensurability of frequencies} proponed in \cite{Kry02},\cite{Fon01}: 
{\it no combinations $\omega_l-\omega_{l'}+\omega_{l''}$ can belong to the set 
(1), when all the frequencies are different.}
Now we finished to describe the principle of the network operating. This 
network will be called {\it the parametrical neural network} (PNN).

There are arguments going in for PNN. First of all, the frequency-phase 
modulation is more convenient for optical processing of signals. It 
allows us to back down an artificial adaptation of an optical network to 
amplitude modulated signals. Second, when signals with $q$ different 
frequencies can propagate along one interconnection (this is an analog 
of the channel multiplexing), this, in fact, allows us to reduce the 
number of interconnections by a factor of $q^2$.
Note, interconnections occupy nearly 98\% of the area of neurochips. 
Third, the signal-noise analysis made with the aid of the 
Chebyshev-Chernov statistical method showed that the storage capacity of PNN 
was approximately $q^2$ times as much as the storage capacity of the 
Hopfield model. Even if $q\sim 10$, the gain is two orders. For computer 
processing of colored images the standard value is
$q=256$. Consequently, comparing with the Hopfield model the gain is 
about five orders. Simultaneously with an increase of the storage 
capacity, the noise immunity of the network also increases. For example, 
we simulated PNN with the following parameters: $N=100$, $q=22$ and 
$p=200$. This network recognized any 80\% noisy pattern after 100 steps 
(in fact, in one pass over all the neurons).
The same network with the parameters $N=100,\ q=25,\ p=1000(!)$ 
recognized a 65\% noisy pattern in 4-5 passes over all the neurons. We 
remind that some time ago the ability of the Hopfield model with 
$N=400,\ p=30$ to recognize a 30\% noisy pattern was 
presented as a high mark in the patterns recognition \cite{Kin87}. 

In the present work we investigate the abilities of PNN. Here an 
important remark has to be done. Generally speaking, there are different 
parametrical FWM-processes complying with the principle of incommensurability of 
frequencies. For example in \cite{Kry02},\cite{Fon01} the parametrical 
FWM-process of the type
$$\omega_l-\omega_{l'}+\omega_{l''}=\left\{ 
\begin{array}{cl}\omega_{l''},&\mbox{ when }l'=l;\\
\omega_l,&\mbox{ when }l'=l'';\\
\rightarrow 0&\mbox{ in other cases.}\end{array}\right.
$$
was examined. The corresponding network will be called PNN-I. However,
better results can be obtained for the parametrical FWM-process  
$$\omega_l-\omega_{l'}+\omega_{l''}=\left\{
\begin{array}{cl}
\omega_l,&
\mbox{ when}l'=l'';\\
\rightarrow 0&\mbox{in other cases.}\end{array}
\right.\eqno(4)$$
This network will be called PNN-II.

The organization of the paper is as follows. In Section 2 we introduce a 
vector formalism allowing us to formulate the problem in the general 
form. In this section the results for PNN-II are presented. In Section 3 
the vector formalism is used to examine the Potts-glass neural network. 
We compare it with PNN-II. Some remarks are given in Conclusions. The 
details of calculations are in Appendix.

\section{PNN-II}
In fact, PNN is an associative memory of the Hopfield type with neurons, 
which can be in more than two different states. Such models of 
neural networks were examined previously (see, for example, \cite{Noe}--
\cite{Bol2}). Usually neurons are modeled with the aid 
of vectors, but not scalar quantities equal to $\pm 1$ or $0/1$. The 
number of {\it the representative vectors} is equal to the number of different 
states of neurons. In the present Section for PNN-II we formulate the 
vector formalism and then estimate the storage capacity of such a 
network.

\subsection{Vector Formalism}
In order to describe the $q$ different states (1) of neurons we use the 
set of basis vectors ${\bf e}_l$ in the space $\rm R^q,\ q\ge 1$,
$${\bf e}_l=\left(\begin{array}{c}0\\\vdots\\1\\
\vdots\\0\end{array}\right),\ l=1,\ldots,q.$$
The state of the $i$th neuron is described by a vector 
${\bf x}_i$,
$${\bf x}_i=x_i{\bf e}_{l_i},\quad x_i=\pm 1,\ {\bf e}_{l_i}\in{\rm 
R^q},\ 
1\le l_i\le q,\ i=1,\ldots,N.$$ 
The state of the network as a whole $X$ is determined by a set of $N$ 
$q$-dimensional vectors ${\bf x}_i$: $X=({\bf x}_1,\ldots,{\bf x}_N)$.
By analogy with Eq.(2) the $p$ stored patterns are 
$$X^{(\mu)}=({\bf x}^{(\mu)}_1,{\bf x}^{(\mu)}_2,\ldots,
{\bf x}^{(\mu)}_N),\quad
{\bf x}^{(\mu)}_i=x^{(\mu)}_i{\bf e}_{l^{(\mu)}_i},\ x^{(\mu)}_i=\pm 1,
\ \left\{\begin{array}{l}1\le l^{(\mu)}_i\le q,\\
\mu=1,\ldots,p.\end{array}\right.$$ 
Since in this model neurons are vectors, the local field 
${\bf h}_i$ affecting the $i$th neuron is a vector too. By analogy with 
the standard Hopfield model we write
$${\bf h}_i  = \sum\limits_{j = 1}^N {\bf T}_{ij}{\bf x}_j\eqno(5)$$
The $(q \times q)$-matrix ${\bf T}_{ij}$ describes the interconnection 
between the $i$th and the $j$th neurons. This matrix affects the vector 
${\bf x}_j\in\rm R^q$, converting it in a linear combination of basis 
vectors ${\bf e}_l$. This combination is an analog of the packet of 
quasi-monochromatic pulses that come from the $j$th neuron to the $i$th 
one after transformation in the interconnection (see Introduction). To 
satisfy the conditions (3) and (4), we need to take the matrices ${\bf 
T}_{ij}$ as
$${\bf T}_{ij}=(1-\delta_{ij})\sum_{\mu=1}^p{\bf x}^{(\mu)}_i
{{\bf x}_j^{(\mu)}}^+,\quad i,j=1,\ldots,N,\eqno(6)$$
where $\delta_{ij}$ is the Kronecker symbol. The elements of these 
matrices are
$$T_{ij}^{(kl)}=
(1-\delta_{ij})\sum\limits_{\mu=1}^p({\bf e}_k{\bf x}^{(\mu)}_i)
({\bf x}^{(\mu)}_j{\bf e}_l)
\quad k,l=1,\ldots,q.
$$
Let us define the dynamics of our $q$-dimensional neurons. Let 
$X(t)=({\bf x}_1(t),\ldots,{\bf x}_N(t))$ be the state of the system at 
the time $t$. By definition the $i$th neuron at the time $t+1$ is 
oriented along a direction mostly close to the local field ${\bf 
h}_i(t)$. Let us clarify this definition. With the aid of (6) we 
write Eq.(5) in the form more convenient for analysis:
$${\bf h}_i(t)=\sum_{l=1}^q A^{(i)}_l
{\bf e}_l,
\mbox{ where }
A^{(i)}_l=\sum\limits_{j(\ne i)}^N\sum\limits_{\mu=1}^p({\bf e}_l
{\bf x}^{(\mu)}_i)({\bf x}^{(\mu)}_j{\bf x}_j(t)).\eqno(7)$$
Let $k$ be the index relating to the amplitude that is maximal in 
modulus in the series (7): 
$$\begin{array}{rcl}\mid A^{(i)}_k\mid=&\max&\mid A^{(i)}_l\mid.\\
{}&1\le l\le q&{}\end{array}$$
Then according to our definition 
$${\bf x}_i(t+1)={\rm sgn}(A^{(i)}_k){\bf e}_k.\eqno(8)$$
The expression (8) is identical to the 'winner-take-all' rule of Introduction.

The evolution the system consists of consequent changes of orientations 
of vector-neurons according to the rule (8). We make the convention that if 
some of the amplitudes are maximal in modulus simultaneously, and the neuron is 
in one of these {\it unimprovable} states, its state does not change. 
Then it is easy to show that during the evolution of the network its
{\it energy} $H(t)=-1/2\sum_{i=1}^N({\bf h}_i(t){\bf x}_i(t))$ decreases.
In the end the system reaches a local energy minimum. In this 
state all the neurons ${\bf x}_i$ are oriented in an unimprovable 
manner, and the evolution of the system come to its end. These states 
are the fixed points of the system. The necessary and sufficient 
conditions for a configuration $X$ to be a fixed point is fulfillment of 
the set of inequalities:
$$({\bf x}_i{\bf h}_i)\ge \mid ({\bf e}_l{\bf h}_i)\mid,\quad 
\forall\ l=1,\ldots,q;\ \forall\ i=1,\ldots,N.\eqno(9)$$

\subsection{Storage capacity of PNN-II}
Let us estimate the storage capacity of the network in the limit $N\gg 
1$. 
Suppose that the network starts from a distorted $m$th pattern 
$$\tilde X^{(m)}=(a_1\hat b_1{\bf x}^{(m)}_1, a_2\hat b_2{\bf 
x}_2^{(m)},
\ldots,a_N\hat b_N{\bf x}_N^{(m)}).$$
Here $\{a_i\}_1^N$ and $\{\hat b_i\}_1^N$ define a {\it phase noise} and a 
{\it frequency noise}
respectively: $a_i$ is a random value that is equal to 
$-1$ or $+1$ with the probabilities $a$ and $1-a$ respectively; $b$ is 
the probability that the operator $\hat b_i$ changes the state of the 
vector ${\bf x}^{(m)}_i=x^{(\mu)}_i{\bf e}_{l^{(\mu)}_i}$ , and $1 - b$ 
is the probability that this vector remains unchanged.

Let us examine to what extent the neural network recognizes the pattern 
$X^{(m)}$ correctly. The amplitudes $A^{(i)}_l$ (7) have the form 
$$A^{(i)}_l=\left\{\begin{array}{rcll}
x^{(m)}_i\sum_{j=1}^{N-1}\xi_j&+&
\sum_{r=1}^L\eta_r(l^{(m)}_i),&\mbox{ when }l=l^{(m)}_i;\\
{}&{}&\sum_{r=1}^L\eta_r(l),&\mbox{ when }l\ne l^{(m)}_i,
\end{array}\right.\eqno(10)$$
where $\xi_j=a_j({\bf x}^{(m)}_j\hat b_j{\bf x}^{(m)}_j)$, 
$\eta_r(l)\equiv\eta_j^{(\mu)}(l)=a_j({\bf e}_l{\bf x}^{(\mu)}_i)
({\bf x}^{(\mu)}_j\hat b_j{\bf x}^{(m)}_j)$, $j=1,\ldots,N$, $j\ne i$,
$\mu=1,\ldots,p$, $\mu\ne m$. For simplicity, when writing the quantities 
$\eta$ in place of the superscript $\mu$ and the subscript $j$ we use 
the subscript $r$ which takes $L=(N-1)(p-1)$ different values
$r=1,\ldots,L$.

Let us note that when the patterns  
$\{X^{(\mu)}\}_1^p$ are uncorrelated, the quantities  
$\xi_j$ and $\eta_r$ can be considered as independent random variables 
described by the probability distributions
$$\xi_j=\left\{ {\begin{array}{*{20}c}
   { + 1,}  \\
   {0,}  \\
   { - 1}  \\
\end{array}} \right.\begin{array}{*{20}c}
   {(1 - a)(1 - b)} \hfill  \\
   b \hfill  \\
   {(1 - a)b} \hfill  \\
\end{array}
,\quad
\eta_r(l)=\left\{{\begin{array}{*{20}c}
   { + 1,}  \\
   {0,}  \\
   { - 1}  \\
\end{array}} \right.\begin{array}{*{20}c}
   {1/2q^2 } \hfill  \\
   {1 - 1/q^2 } \hfill  \\
   {1/2q^2 } \hfill  \\
\end{array}.\eqno(11)
$$
Since the distributions of the quantities $\eta_r(l)$ are independent of 
$l$, in what follows we simply write $\eta_r$. According to the rule 
(9), the $i$th neuron finds itself in the state ${\bf x}^{(m)}_i$ when 
two conditions for the amplitudes (10) are fulfilled:
$${\rm sgn}(A^{(i)}_{l_i^{(m)}})=x_i^{(m)},\quad 
\sum_{j=1}^{N-1}\xi_j + x^{(m)}_i
\sum_{r=1}^L\eta_r\ge\mid\sum_{r=1}^L\eta_r\mid.$$
Otherwise there will be an error in the recognition of the vector 
${\bf x}^{(m)}_i $. Since the random variable $x^{(m)}_i\eta_r$ has the 
same distribution as $\eta_r$, the probability of this error is
$${\Pr}_i=\Pr\left\{\sum_{j=1}^{N-1}\xi_j+\sum_{r=1}^L\eta_r<0\right\}.
\eqno(12)$$
To estimate the value of ${\Pr}_i$ we use the well-known 
Chebyshev-Chernov method \cite{Che52},\cite{Kry00} (see Appendix). 
As a result we obtain the expression for the probability of the 
error in the recognition of the pattern $X^{(m)}$: 
$${\Pr}_{err}= N\exp\left(-\frac{N(1 - 2a)^2}{2p}\cdot q^2(1-
b)^2\right)\eqno(13)$$
When $N$ increases, this probability tends to zero, if $p$ as function 
of $N$ increases slower than
$$p_c=\frac{N(1-2a)^2}{2\ln N}\cdot q^2(1 - b)^2\eqno(14)$$
This allows us to use (14) as an asymptotically possible value of the 
storage capacity of PNN-II.

When $q = 1$, Eqs.(13)-(14) transform into 
well-known results for the standard Hopfield model
(in this case there is no frequency noise, $b=0$). When $q$ increases, 
the probability of the error (13) decreases exponentially, i.e. the
noise immunity of PNN increases noticeably. In the same time the storage 
capacity of the network increases proportionally 
to $q^2$. In contrast to the Hopfield model the number of the 
patterns $p$ can be much greater than the number of neurons. 

For example, let us set a constant value ${\Pr}_{err}=0.01$. In the Hopfield 
model, with this probability of the error we can recognize any of $p=N/10$ 
patterns, each of which is less then 30\% noisy. In the same time,  PNN-II with 
$q=64$ allows us to recognize any of $p=5N$ patterns with 90\% noise, or 
any of $p=50N$ patterns with 65\% noise.

In Fig.1 we give an example of the restoration of 90\% distorted pattern ($a=0,
\ b=0.9$). Here the parameters of the network are $N=100$, $p=200$, $q=32$.
The pattern is a picture of a dog. The gray squares are noisy pixels.  
The states of the network after 50 and 100 steps are shown. 

In Fig.2 for different values of $q$ we show the dependence of the probability 
of a pattern recognition ${\rm\bf P}_{\rm rec}=1-{\Pr}_{err}$ as function of the 
frequency noise ${\rm\bf b}=b\cdot 100\%,\ b\in [0,1]$, when $\alpha=p/N=2$ 
(solid line); the phase noise is equal to zero, $a=0$. We see that if $q=20$,
we can recognize correctly any pattern when the noise is less than 70\%, 
and if $q=30$, any pattern when the noise is less than 85\%. Generally,
if the noise is less than a critical value $b_c$, 
$$b_c=1-\frac{2}{q}\sqrt{\frac{p}{N}},\eqno(15)$$
PNN can recognize a noisy 
pattern for sure, and if $b>b_c$, the probability of recognition tends to zero.
Our computer simulations confirm these results.

\section{Potts-glass neural networks}
The models of associative memory with neurons that can be in more than two 
different states have been investigated by a lot of authors \cite{Noe}--
\cite{Bol2}. All these models are related with the Potts model of magnetic. The 
last generalizes the Ising model for the case of the spin variable that takes 
$q>2$ different values \cite{Wu},\cite{Bax}. In all these works the authors used 
the same well-known approach relating the Ising model with the Hopfield model 
(see, for example, \cite{Her}). Namely, in place of the short-range interaction 
between two nearest spins the Hebb type interconnections between all 
vector-neurons 
were used. As a result, long-range interactions appeared. Then in the mean-field 
approximation it was possible to calculate the statistical sum and, 
consequently, to construct the phase diagram. Different regions of the phase 
diagram were interpreted in the terms of the ability of the network to recognize 
noisy patterns.

Among all the models of $q$-state associative memory, characteristics of {\it 
the anisotropic Potts-glass neural network} (APGNN) \cite{Kan88}, 
\cite{Bol1},\cite{Bol2} are most close to PNN-II. In other models the storage 
capacity is less than even for the Hopfield 
model. Below we describe APGNN in terms of our vector formalism and compare it 
with PNN-II.

APGNN consists of $N$ neurons each of which can be in $q$ different states. Now 
to describe the states of the neurons in place of the basis vectors ${\bf 
e}_l\in{\rm R^q}$ (see Subsection 2.1) $q$-dimensional vectors of a special type 
are used. Namely, the $l$th state of a neuron is described by a column-vector 
${\bf d}_l\in{\rm R^q}$,$${\bf d}_l=\frac1q\left(\begin{array}{c}-1\\\vdots\\q-
1\\
\vdots\\-1\end{array}\right),\ l=1,\ldots,q.$$
The state of the $i$-th neuron is described 
by a vector ${\bf x}_i={\bf d}_{l_i}$, $1\le l_i\le q$, $i=1,\ldots,N$.
The state of the network as a whole $X$ is determined by a set of $N$ 
$q$-dimensional vectors ${\bf x}_i$: $X=({\bf x}_1,\ldots,{\bf x}_N)$.
The $p$ stored patterns are 
$$X^{(\mu)}=({\bf x}^{(\mu)}_1,\ldots,
{\bf x}^{(\mu)}_N),\quad 
{\bf x}^{(\mu)}_i={\bf d}_{l^{(\mu)}_i},\ 1\le l^{(\mu)}_i\le q,
\ \mu=1,2,\ldots,p.$$ 
The local field ${\bf h}_i$ affecting the $i$-th neuron is the vector
${\bf h}_i  = \sum\limits_{j = 1}^N {\bf T}_{ij}{\bf x}_j$, 
where $(q \times q)$-matrices ${\bf T}_{ij}$ describe the 
interconnections between the $i$-th and the $j$-th neurons. 
These matrices are
$${\bf T}_{ij}=(1-\delta_{ij})\sum_{\mu=1}^p{\bf x}^{(\mu)}_i
{{\bf x}_j^{(\mu)}}^+,\quad i,j=1,\ldots,N.$$
The same as in Subsection 2.1 the dynamics of APGNN is defined as follows:
the $i$-th neuron at the next time step $t+1$ is oriented 
along a direction mostly close to the local field ${\bf h}_i(t)$ at the time 
$t$. 
During the evolution of the network the energy $H(t)=-1/2\sum_{i=1}^N({\bf 
h}_i(t){\bf x}_i(t))$ decreases. The necessary and sufficient 
conditions for a configuration $X$ to be a fixed point is fulfillment of 
the set of inequalities:
$$({\bf x}_i{\bf h}_i)\ge ({\bf d}_l{\bf h}_i),\quad 
\forall\ l=1,\ldots,q;\ \forall\ i=1,\ldots,N.$$

We see that PNN-II and APGNN are much alike. The difference between these models 
is that, first, in APGNN the vectors ${\bf d}_l$ are 
nonorthogonal, and, second, in APGNN there are no amplitudes $\pm 1$ relating 
with the vectors ${\bf d}_l$.

When $q=2$, APGNN is the same as the standard Hopfield model\cite{Kan88}. 
Repeating the argumentation of Subsection 2.2. we can estimate the storage 
capacity of APGNN for $N\gg 1$. We must only take into account that there is no 
phase noise $\{a_i\}_1^N$ in this model. The distorted $m$th pattern 
has a form
$$\tilde X^{(m)}=(\hat b_1{\bf x}^{(m)}_1,\hat b_2{\bf x}_2^{(m)},
\ldots,\hat b_N{\bf x}_N^{(m)}).$$
As above, the random operator $\hat b_i$ with the probability $b$ 
changes the state of the vector ${\bf x}^{(m)}_i$, and with the probability 
$1-b$ this vector remains unchanged. Now, the probability of the error in the
recognition of the vector ${\bf x}^{(m)}_i$ is
$${\Pr}_i=\Pr\left\{\sum_{j=1}^{N-1}\xi_j+\sum_{r=1}^L\eta_r(l_I^{(m)})
<\sum_{r=1}^L\eta_r(l)\right\},$$
where the independent random variables 
$\xi_j=({\bf x}^{(m)}_j\hat b_j{\bf x}^{(m)}_j)$, 
$\eta_r=({\bf d}_l{\bf x}^{(\mu)}_i)({\bf x}^{(\mu)}_j
\hat b_j{\bf x}^{(m)}_j)$ are distributed
$$\xi_j=\left\{\begin{array}{cc}(q-1)/q,&1-b\\-1/q,&b\end{array}\right.,
\quad
\eta_r=\left\{\begin{array}{cc}(q-1)/q,&1/q^2\\
1/q,&(q-1)/q^2\\
0,&(q-2)/q\\
-1/q,&(q-1)/q^2\\
-(q-1)/q,&1/q^2\end{array}\right.$$
Naturally, it is true for the randomized patterns $\{X^{(\mu)}\}_1^p$, only.

Similarly to calculations of Appendix we obtain the expression for the 
probability of the error in the recognition of the pattern $X^{(m)}$, 
$${\Pr}_{err}=N\exp\left(-\frac{N}{2p}\frac{q(q-1)}2(1-\bar b)^2\right),
\quad\bar b=\frac{q}{q-1}b.\eqno(16)$$ 
Then the asymptotically possible value of the storage capacity of APGNN is
$$p_c=\frac{N}{2\ln N}\frac{q(q-1)}2(1-\bar b)^2.\eqno(17)$$ 
When $q=2$, these expressions give the known estimates for the Hopfield model.
For $q>2$ the storage capacity of APGNN is $q(q-1)/2$ times as large as the 
storage capacity of the Hopfield model. In \cite{Kan88} the same factor was 
obtained by fitting the results of numerical calculations. Our approach allows 
us to obtain the same result rigorously.

\section{Conclusions}
For $q>>1$ the storage capacity of APGNN is two times less than the storage 
capacity of PNN-II (compare Eq.(17) with Eq.(14) for $a\sim 0$). When 
calculating the probability of the error in the recognition,
the additional factor two appears in the exponent (see Eqs.(13),(16)). This 
leads to a significant decrease of a noise immunity of APGNN comparing with 
PNN-II. This is well seen in Fig.2, where
for APGNN the dashed line shows the dependence of the pattern recognition 
${\rm\bf P}_{\rm rec}$ on the value of the frequency noise $\rm\bf b$ under the
same conditions as for PNN-II (the solid line). The superiority of PNN-II is 
easily seen, especialy in the region of not so large values of $q\sim 10$. For 
APGNN the critical value of the noise $b_c$ (15) is less than the
analogous characteristic of PNN-II by a quantity 
$\frac{0.8}{q}\sqrt{\frac{p}{N}}$.

In conclusion, we would like to note that our approach allows 
us to describe not only the optical neural networks of the parametrical type, 
but also neural networks in which information is encoded in the form of  
phase delays of pulses in interconnections. It is much more easy to realize 
such a network in form of a device.

\section{Appendix}
The following equation is true for the probability ${\Pr}_i$ (12):
 $${\Pr}_i\le 
\Pr\left\{\sum_{j=1}^{N-1}\xi_j+\sum_{r=1}^L\eta_r\le 0\right\}=
\Pr\left\{-\sum_{j=1}^{N-1}\xi_j-\sum_{r=1}^L\eta_r\ge 0\right\}
$$
Using the known approach of exponential estimates of the Chebyshev type, for any 
positive $z>0$ we obtain:
$${\Pr}_i\le
\overline{\exp\left(z\left(-\sum_{j=1}^{N-1}\xi_j-
\sum_{r=1}^L\eta_r\right)
\right)}=
\left(\overline{\exp(-z\xi_j)}
\left(\overline{\exp(-z\eta_r)}\right)^{p-1}\right)^{N-1}
$$
The over-line means an averaging over all possible realizations, and the last 
equality follows from independence of the random variables $\xi_j$ and $\eta_r$.

Taking into account the distributions (11), it is easy to  obtain
the averages
$$\overline{\exp(-z\xi_j)}=(1-a)(1-b)e^{-z}+b+a(1-b)e^z,\ 
\overline{\exp(-z\eta_r)}=e^{-z}/2q^2+1-1/q^2+e^z/2q^2.$$
Changing the variables $e^z=x$ and introducing functions $f_1(x)$ and
$f_2(x)$,
$$f_1(x)=a(1-b)x+b+\frac{(1-a)(1-b)}x,\ 
f_2(x)=\frac1{2q^2}\left(x+\frac1x\right)
+1-\frac1{q^2},$$
we obtain that for any positive $x$ the following estimate is valid: 
$${\Pr}_i\le \left(f_1(x)f_2^{p-1}(x)\right)^{N-1}.\eqno(18)$$
To obtain the minimal possible value of the probability ${\Pr}_i$, we need to 
find the value of the variable $x$ minimizing the right-hand side of Eq.(18). 
This leads us to the equation
$$(p-1)(x^2-1)+
\frac{a(1-b)x^2-(1-a)(1-b)}{a(1-b)x^2+bx+(1-a)(1-b)}(x^2+2(q^2-
1)x+1)=0.$$
When $p\gg 1$, the proper root of this equation up to the terms of the order of 
$1/p$ is equal to $x_1=1+q^2(1-2a)(1-b)/(p-1)$. Substituting this value of $x$ 
in Eq.(18), we obtain 
$${\Pr}_i\le \left(1-\frac{q^2(1-2a)^2(1-b)^2}{2(p-1)}\right)^{N-1}\cong
\exp\left(-\frac{N(1-2a)^2}{2p}\cdot q^2(1-b)^2\right).$$
This inequality gives the estimate (13) for the probability of the error in the 
recognition of the pattern $X^{(m)}$.
\vskip 10mm
\centerline{\bf List of Figures}
\vskip 2mm

{\bf Fig.1.} The restoration of the pattern with 90\% frequency noise 
($b=0.9$), when $N=100$, $p=200$ and $q=32$. The pattern is a picture of a
dog. The gray squares are noisy pixels. The states of the network after 50 and
100 steps are shown.  

{\bf Fig.2.} The probability of the pattern recognition 
${\rm\bf P}_{\rm rec}=1-{\Pr}_{err}$ versus frequency noise 
${\rm\bf b}=b\cdot 100\%,\ b\in [0,1]$ for different values of $q$ and 
$\alpha=p/N=2$ for PNN-II (solid line) and for the Potts-glass neural
network (dashed line).

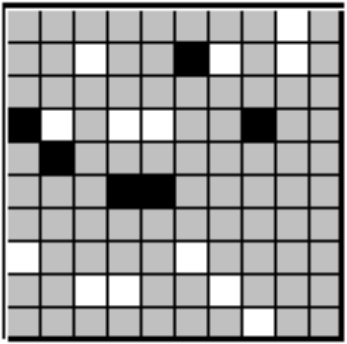
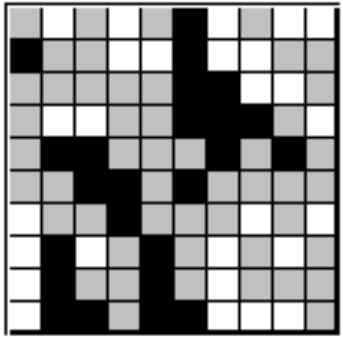
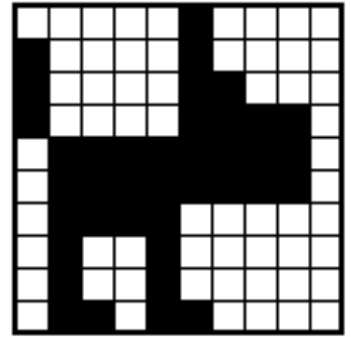
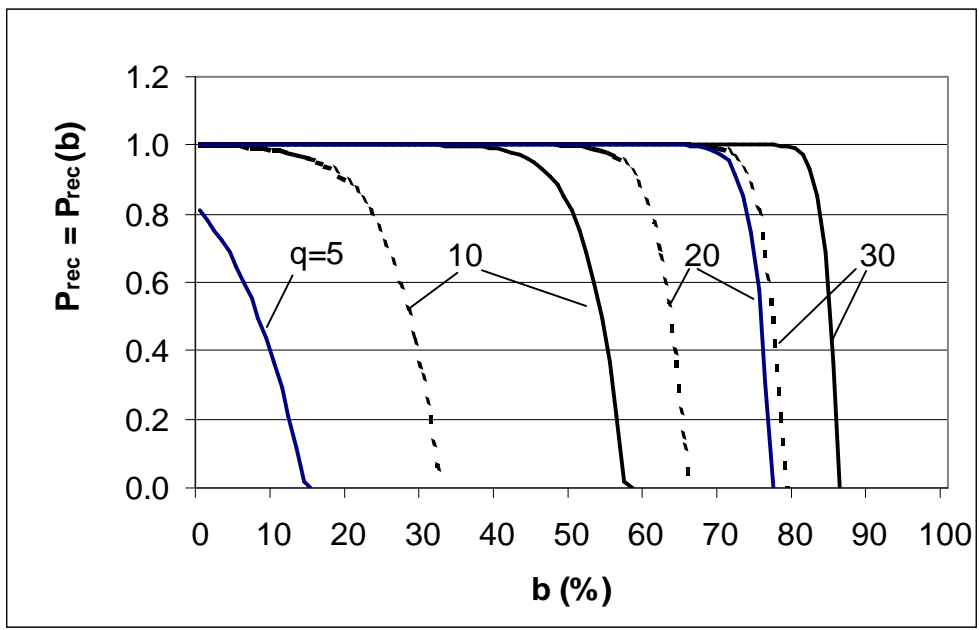
\end{document}